\documentclass[twocolumn,showpacs,preprintnumbers,amsmath,amssymb]{revtex4-1}

\usepackage{amssymb}
\usepackage{amsmath}
\usepackage{graphicx}
\usepackage{color}
\usepackage{hyperref}
\usepackage{slashbox}

\newcommand{\be}{\begin{equation}}
\newcommand{\ee}{\end{equation}}
\newcommand{\bea}{\begin{eqnarray}}
\newcommand{\eea}{\end{eqnarray}}
\newcommand{\beal}{\begin{aligned}}
\newcommand{\eeal}{\end{aligned}}

\usepackage{color}

\begin{document}



\title{Black hole evaporation in Lovelock gravity with diverse dimensions}


\author{Hao Xu}
\email{xuh5@sustc.edu.cn}
\affiliation{Institute for Quantum Science and Engineering, Department of Physics, Southern University of Science and Technology, Shenzhen 518055, China}
\affiliation{Department of Physics, University of Science and Technology of China, Hefei 230026, China}

\author{Man-Hong Yung}
\affiliation{Institute for Quantum Science and Engineering and Department of Physics, Southern University of Science and Technology, Shenzhen 518055, China}
\affiliation{Shenzhen Key Laboratory of Quantum Science and Engineering, Shenzhen, 518055, China}


\date{\today}

\begin{abstract}
We investigate the black hole evaporation process in Lovelock gravity with diverse dimensions.
By selecting the appropriate coefficients, the space-time solutions can possess a unique AdS
vacuum with a fixed cosmological constant $\Lambda=-\frac{(d-1)(d-2)}{2\ell^{2}}$.
The black hole solutions can be divided into two cases: $d>2k+1$ and $d=2k+1$. In the case of
$d>2k+1$, the black hole is in an analogy with the Schwarzschild AdS black hole, and
the life time is bounded by a time of the order of $\ell^{d-2k+1}$, which reduces Page's result
on the Einstein gravity in $k=1$. In the case of $d=2k+1$, the black hole resembles the
three dimensional black hole. The black hole vacuum corresponds to $T=0$, so the black hole
will take infinite time to evaporate away for any initial states, which obeys the third law
of thermodynamics. In the asymptotically flat limit $\ell\rightarrow \infty$, the system reduces
to the pure Lovelock gravity that only possesses the highest $k$-th order term. For an initial
mass $M_0$, the life time of the black hole is in the order of $M_0^{\frac{d-2k+1}{d-2k-1}}$.

\end{abstract}


\maketitle

\section{Introduction}

In modern gravity theories, the effects of gravitation are ascribed to space-time curvature instead of a force. The gravity equations relate the presence of matter and the curvature of space-time. Black holes, which were first discovered as a particular class of solutions of the Einstein gravity and considered as a mathematical curiosity, are in fact inevitable under quite generic initial conditions in almost any reasonable gravity theories. Schwarzschild black hole, according to Birkhoff's theorem, is the most general static spherically symmetric vacuum solution of the Einstein gravity. It also describes the leading asymptotic behavior of the geometry for any localized distribution of matter in perturbation theory.

By applying quantum field theory in curved space-time, Hawking showed the quantum effects allow black holes to emit particles, known as Hawking radiation \cite{Hawking:1974rv,Hawking:1974sw}. Insight into the process may be acquired by imagining particle-antiparticle pair creation near the event horizon. One of the pair falls into the black hole while the other escapes. To an outside observer, it would appear that the black hole has just emitted a particle. Hawking radiation connects gravity theories and quantum field theory, and it is expected to shed light on quantum gravity.

If there is no incoming matter to balance the Hawking radiation, the black hole continues to lose mass and thermal entropy. For a Schwarzschild black hole with initial mass $M_0$ in four-dimensional asymptotically flat space-time in Einstein gravity, we expect it has a life time $t\sim M_0^3$, which is divergent when $M_0\rightarrow\infty$ \cite{Page:1976df,Page:2004xp}. Similar analysis can also be carried out for a spherical black hole in AdS space-time, if an absorbing AdS boundary condition is chosen so that the radiation particles cannot reflect back \cite{Avis:1977yn}. In \cite{Page:2015rxa}, Page made a counter-intuitive discovery that the life time of black hole in 4-dimensional AdS space-time does not diverge even when the initial black hole mass is taken to be infinity. In fact, it is bounded by a time of the order of $\ell^3$, where $\ell$ is the AdS radius.

On the other hand, higher derivative gravity theories have attracted a considerable amount of interest as an alternative theory to Einstein gravity. Partly due to the AdS/CFT correspondence \cite{Maldacena:1997re,Witten:1998qj}, the higher derivative terms may serve as the corrections to the large $N$ expansion in the dual conformal field theory, and partly due to the inconsistency between the prediction of the Einstein gravity and observation of galactic rotation curves, the higher derivative gravity, such as conformal (Weyl) gravity, may produce the effective potential consistent with the observed phenomenon at large scale, without introducing the unknown ※dark matter§, while maintaining the correct behaviors at the scale of the Solar system \cite{Mannheim:1988dj,Mannheim:2005bfa,Mannheim:2010ti,Mannheim:2011ds}.

There are also their own inherent defects in higher derivative gravity theories. The higher powers of curvature could give rise to higher order differential equations for the metric, which would introduce ghosts and violate unitarity. However, one of the higher derivative gravity theories, known as Lovelock gravity \cite{Lovelock:1971yv}, could overcome the problem. Its action contains a sum of dimensionally-extended Euler densities, so that the integral of the $k$-th order term in Lovelock gravity gives the Euler character in dimension $d=2k$. When $d<2k$, the $k$-th order term vanishes identically, hence, it contributes to the space-time solution only in dimensions $d\geq 2k+1$. The equations of motion, which depend only on the Riemann tensor and not on its derivatives, remain second order. In fact, the kinetic term of Einstein-Hilbert action and cosmological constant are precisely the first-order and zeroth-order terms respectively. Lovelock gravity arises as a natural generalization of Einstein gravity, and its physics properties have been studied in lots of literatures \cite{Kastor:2010gq,Cai:2013qga,Xu:2013zea,Xu:2014tja,Wei:2014hba,Frassino:2014pha,Dolan:2014vba,Xu:2015hba,Konoplya:2017ymp,Konoplya:2017lhs,Toledo:2019szg,Xu:2019xif,Ovgun:2019ygw}.

However, beyond the perturbation theory, the higher powers of curvature make the Lovelock gravity different from the Einstein gravity. If the coefficients before the Euler density terms are chosen arbitrarily, the Lovelock gravity can have some drawbacks. The dynamical evolution can become unpredictable and the Hamiltonian is ill defined \cite{Teitelboim:1987zz,Henneaux:1987zz}. The space-time solutions may admit negative energy solutions with event horizons or positive energy solutions with naked singularities \cite{Wheeler:1985nh,Cai:2006pq}.

These problems can be overcome by selecting the appropriate coefficients, demanding the space-time solutions possess a unique AdS vacuum with a fixed cosmological constant \cite{Crisostomo:2000bb,Aros:2000ij}. The aim of the present work is to investigate the black hole evaporation process in this situation. We adopt the units of \cite{Page:2015rxa}, using $\hbar=c=k_{\textrm{Boltzmann}}=1$.

\section{Black hole thermodynamics in Lovelock gravity}

We start by giving a brief review of the black hole thermodynamics in Lovelock gravity. Detailed analysis including topologically nontrivial AdS asymptotics and electric charge can be found in \cite{Crisostomo:2000bb,Aros:2000ij}. In the present work we focus on the simplest class of the theory, the static spherical black hole with unique negative cosmological constant.

The action of the theory can be written as \cite{Lovelock:1971yv}
\begin{equation}
I_{k}=\kappa \int \sum_{p=0}^{k}c_{p}^{k}L^{(p)}\;,  \label{Ik}
\end{equation}
where $L^{(p)}$ represents the $p$-th order Euler density, and the integer $k$, which satisfies $1\leq k\leq [\frac{d-1}{2}]$, represents the highest power of curvature in the action. The Lovelock coefficients $c_{p}^{k}$ are chosen to be \cite{Crisostomo:2000bb}
\begin{equation}
c_{p}^{k}=\left\{
\begin{array}{ll}
\frac{\ell^{2(p-k)}}{(d-2p)}\left(
\begin{array}{c}
k \\
p
\end{array}
\right)  & ,\text{ }p\leq k \\
0 & .\text{ }p>k
\end{array}
\right.
\label{Coefs}
\end{equation}
The two constant $\kappa$ and $\ell$, are related with the gravitational constant $G_k$ and the negative cosmological constant $\Lambda$ as
\begin{eqnarray}
\kappa &=&\frac{1}{2(d-2)!\Omega _{d-2}G_{k}},  \label{Kappa} \\
\Lambda &=&-\frac{(d-1)(d-2)}{2\ell^{2}}.  \label{Lambda}
\end{eqnarray}

We can obtain the equations of motion by applying the variation method on the above action. If we consider $d$-dimensional static spherical black hole in terms of Schwarzschild-like coordinates, the metric can be written as \cite{Crisostomo:2000bb}
\begin{equation}
\mathrm{d}s^{2}=-f(r)\mathrm{d}t^{2}+\frac{\mathrm{d}r^{2}}{f(r)}+r^{2}\mathrm{d}\Omega
_{d-2}^{2},  \label{gspherical}
\end{equation}
where
\begin{eqnarray}
f(r)=1+\frac{r^{2}}{\ell^{2}}-\left( \frac{2G_{k}(M+C_{0})}{r^{d-2k-1}}%
\right) ^{1/k}.
\end{eqnarray}
Here the $M$ stands for the black hole mass and $C_0$ is a integral constant. The black hole radius $r_+$ satisfies $f(r_+)=0$ so the black hole mass $M$ is written as
\begin{equation}
M(r_{+})=\frac{r_{+}^{d-2k{\bf -}1}}{2G_{k}}\left( 1+\frac{r_{+}^{2}}{\ell^{2}}%
\right) ^{k}-C_{0}.  \label{mass(r)}
\end{equation}
When the black hole radius $r_+$ shrinks to a point, the mass $M\rightarrow 0$, so we can obtain
\begin{equation}
C_{0}=\frac{1}{2G_{k}}\delta _{d-2k,1}\;.  \label{C0}
\end{equation}
Then the $f(r)$ takes the form
\begin{eqnarray}
f(r)=1+\frac{r^{2}}{\ell^{2}}-\bigg(\frac{2G_k M+\delta_{d-2k,1}}{r^{d-2k-1}}\bigg)^{1/k}.
\label{f}
\end{eqnarray}
For $k=1$, the three dimensional black hole \cite{Banados:1992wn} and Schwarzschild AdS black hole in $d$-dimensional space-time can be recovered. One can see that for any value of $k$, $d=2k+1$ differs from other higher dimensions. For $d>2k+1$, the black hole is in an analogy with the Schwarzschild AdS black hole, which has a continuous mass spectrum, whose vacuum state is the pure AdS space-time. However, for $d=2k+1$ the pure AdS space-time, which is obtained for $M=-\frac{1}{2G_k}$, differs from the black hole vacuum $M=0$, predicting the existence of a mass gap separating the pure AdS space-time and black hole vacuum.

The temperature is proportional to the surface gravity at the black hole horizon
\begin{equation}
T=\frac{1}{4\pi}\left.\frac{\mathrm{d}f}{\mathrm{d}r}\right| _{r_{+}}\;=\frac{1}{4\pi k}\left( (d-1)\frac{r_{+}}{\ell^{2}}+\frac{(d-2k-1)}{%
r_{+}}\right).
\label{temperature}
\end{equation}
We can also find the case of $d=2k+1$ is qualitatively different from other choices, since the last term in $T$ vanishes when $d=2k+1$. Consequently, the temperature $T$ vanishes as $r_+\rightarrow 0$ when $d=2k+1$, while it becomes divergent as $r_+\rightarrow 0$ in higher dimensions.

In the case of $d>2k+1$, the black hole is in an analogy with the Schwarzschild AdS black hole of Einstein gravity, which also possesses a Hawking-Page like phase transition \cite{Hawking:1982dh}. The temperature $T$ admits a minimum at
\begin{equation}
r_{c}=\ell\sqrt{\frac{d-2k-1}{d-1}},  \label{rc}
\end{equation}
and the corresponding temperature is
\begin{equation}
T_{c}=\frac{\sqrt{\left( d-2k-1\right) \left( d-1\right) }}{2\pi k\ell},
\end{equation}
while in the case of $d=2k+1$, the black hole temperature $T$ is a monotonically increasing function of $r_+$, and its absolute minimum is at $r_+=0$ and $T=0$.

The black hole entropy, which can be obtained by applying the first law of black hole thermodynamics $\mathrm{d}M=T\mathrm{d}S$ (or calculating partition function through Euclidean path integral in the saddle point approximation \cite{Gibbons:1976ue}), can be written as
\begin{equation}
S_{k}=\frac{2\pi k}{G_{k}}\int_{0}^{r_{+}}r^{(d-2k-1)}\left( 1+\frac{r^{2}}{\ell^{2}}\right) ^{k-1}\mathrm{d}r.
\label{Entropy}
\end{equation}
For the Einstein grivity ($k=1$) in $d=4$, we have
\begin{equation}
S=\frac{\pi r_+^2}{G}=\frac{A}{4G},
\end{equation}
which obeys the celebrated Hawking's area law.\\

Another thing that should be noticed is the asymptotically flat limit $\ell\rightarrow \infty$.
Here we can not simply take the vanishing limit of the zeroth-order term $c_{0}^{k}\rightarrow 0$.
The asymptotically flat limit can only be obtained by setting $\ell\rightarrow \infty$ \cite{Crisostomo:2000bb}, then the
only non-vanishing term in the equation of motion is the
$k$-th one, and the corresponding Lovelock coefficients read
\begin{equation}
c_{p}^{k}=\frac{1}{(d-2k)}\delta _{p}^{k}\;.
\label{Coefs0}
\end{equation}
This system is known as pure Lovelock gravity. We can have the form of $f(r)$ in the black hole metric as
\begin{equation}
f(r)=1-\bigg(\frac{2G_{k}M}{r^{d-2k-1}}\bigg)^{1/k},
\end{equation}
where $d>2k+1$. The case of $d=2k+1$ does not describe black holes. We can also calculate the black hole temperature $T$ and entropy $S$, which read
\begin{equation}
T=\frac{1}{4\pi \kappa _{B}k}\frac{(d-2k-1)}{r_{+}},  \label{Tlim}
\end{equation}
and
\begin{equation}
S_{k}=k\frac{2\pi}{G_{k}}\frac{r_{+}^{(d-2k)}}{(d-2k)}.
\label{SL0}
\end{equation}
When $k=1$, the solution reduces to the Schwarzschild black hole and $f(r)$ has the falloff at large $r$ characteristic of asymptotically flat boundary conditions. The black hole temperature and entropy in Schwarzschild space-time can also be recovered.

\section{Black hole evaporation in Lovelock gravity}
In this section we investigate the black hole evaporation in Lovelock gravity. The black hole mass $M$ should be monotonically-decreasing functions of time. By apply the geometrical optics approximation, which assumes the emitted massless quanta move along null geodesics \cite{Page:2015rxa}, we calculate the life time of the black hole and investigate its relationship with other coefficients. If we orient the extra $(d-3)$ angular coordinates in $\mathrm{d}\Omega_{d-2}^{2}$ and normalize the affine parameter $\lambda$, we have the geodesic equation of the massless quanta reads
\begin{align}
\bigg(\frac{\mathrm{d}r}{\mathrm{d}\lambda}\bigg)^2=E^2-J^2\frac{f(r)}{r^2},
\end{align}
where $E=f(r)\frac{\mathrm{d}t}{\mathrm{d}\lambda}$ and $J=r^2\frac{\mathrm{d}\theta}{\mathrm{d}\lambda}$ are the energy and angular momentum respectively. If the null geodesic of a emitted quanta comes from the black hole horizon, it cannot be detected by the observer on the AdS boundary when there is a turning point $r_0$ ($r_+<r_0<\infty$) satisfying $\big(\frac{\mathrm{d}r}{\mathrm{d}\lambda}\big)^2|_{r_0}=0$.  If we define $b\equiv\frac{J}{E}$, the massless quanta can reach infinity and be observed if and only if
\begin{align}
\frac{1}{b^2}\geq \frac{f(r)}{r^2}
\end{align}
for \emph{any} $r\geq r_+$.

Then our job is to find the maximal value of $\frac{f(r)}{r^2}$, which depends on the exact form of $f(r)$. It may correspond to an unstable photon orbit $r_p$, and the impact factor $b_c=\frac{r_p}{\sqrt{f(r_p)}}$, such as in the spherical AdS black hole case \cite{Page:2015rxa}. Or it may be monotonically increasing function and approaches $\frac{1}{\ell^2}$, so the impact factor $b_c=\ell$, such as in the flat AdS black hole case \cite{Ong:2015fha}.

Once we obtained the impact factor, according to the Stefan-Boltzmann law, we conclude that in $d$-dimensional space-time, the Hawking emission power (luminosity) is
\begin{align}
\frac{\mathrm{d} M}{\mathrm{d}t}=-C b_c^{d-2} T^d.
\label{law}
\end{align}
The above formula is the generalization of the four-dimensional Stefan-Boltzmann law in $d$-dimensional space-time \cite{Vos1988,Vos1989,Cardoso:2005cd}, which implies the emission power should be proportional to the $(d-2)$-dimensional cross section and the photon energy density in $(d-1)$-dimensional space. We can have the photon energy density in $(d-1)$-dimensional space is proportional to $T^d$ by integrating the spectral energy density, and $b_c^{d-2}$ corresponds to the cross section of $d$-dimensional non-rotating black hole \cite{Frolov2011}. Since we are only concerned about the qualitative features of the evaporation process, we shall treat the numerical constant $C$ equal to one. Furthermore, we also set gravitational constant $G_k=1$, meaning the black hole thermodynamic quantities are described in unit of $G_k$, thus we can focus on the black hole evaporation with space-time dimension $d$, cosmological radius $\ell$ and the highest power of curvature $k$. We will investigate them in all different situations.

\subsection{The case of $d>2k+1$}
First of all, we consider the case of $d>2k+1$, which is in an analogy with the well-known Schwarzschild AdS black hole case. The $f(r)$ takes the form
\begin{eqnarray}
f(r)=1+\frac{r^{2}}{\ell^{2}}-\bigg(\frac{2 M}{r^{d-2k-1}}\bigg)^{1/k}.
\label{f}
\end{eqnarray}
Inserting the above formula into $\frac{f(r)}{r^2}$, we can find it possesses a maximal value at the photon orbit $r_p$ reads
\begin{equation}
r_p=\bigg(2M\Big(\frac{d-1}{2k} \Big)^k\bigg)^{\frac{1}{d-2k-1}}.
\end{equation}
When $d=4$ and $k=1$, we have $r_p=3M$, which corresponds to the case of Schwarzschild AdS black hole in 4 dimensions. The impact factor $b_c$ can be obtained as $b_c=\frac{r_p}{\sqrt{f(r_p)}}$.

Inserting the impact factor $b_c=\frac{r_p}{\sqrt{f(r_p)}}$, black hole temperature $T$, and black hole mass $M$ into \eqref{law}, we can have the Hawking emission power and thus investigate the relationship between the evaporation process and initial mass $M_0$, AdS radius $\ell$, highest curvature power $k$, and space-time dimension $d$. However, the exact form of the equation is rather complicated. We introduce $x\equiv \frac{r_+}{\ell}$, then the impact factor $b_c$ and temperature can be rearranged and written as
\begin{equation}
b_c=\ell \frac{x\big(\frac{(1+x^2)(d-1)}{2k}\big)^{\frac{k}{d-2k-1}}}{\bigg(1+x^2\big(\frac{(1+x^2)(d-1)}{2k}\big)^{\frac{2k}{d-2k-1}}-\frac{2k}{d-1}\bigg)^{\frac{1}{2}}},
\end{equation}
and
\begin{equation}
T=\ell^{-1}\frac{1}{4\pi k}\big((d-1)x+\frac{d-2k-1}{x}\big).
\end{equation}
Using the exact form of black hole mass $M$, we also have
\begin{align}
\frac{\mathrm{d} M}{\mathrm{d}t}&=-\ell^{d-2k-1}(1+x^2)^{k-1}\bigg(\frac{d-2k-1}{2}(1+\frac{1}{x^2})+k\bigg)\nonumber\\
               &\times x^{d-2k+2}\frac{\mathrm{d} }{\mathrm{d}t}\bigg(\frac{1}{x}\bigg).
\end{align}
Inserting the above three formulas into the Stefan-Boltzmann law \eqref{law}, we have
\begin{align}
\mathrm{d}t=\ell^{d-2k+1}F(x,d,k)\mathrm{d}\bigg(\frac{1}{x}\bigg),
\end{align}
where $F(x,d,k)$ is a complicated function that is not worth quoting. Integrating the above formula from $\infty$ to $0$, we can find the $\int^0_{\infty} F(x,d,k)\mathrm{d}\big(\frac{1}{x}\big)$ is convergent and depends only on the value of $d$ and $k$. Thus, We can conclude that \emph{the life time of black hole does not diverge with any finite $\ell$, and it remains bounded by a time of the order of $\ell^{d-2k+1}$}. When $k=1$, we can recover Page's result of the Einstein gravity, which states that the total decay time of arbitrarily large black holes in asymptotically AdS space-time has the form $t\sim \ell^{d-1}$.

\begin{figure}
\begin{center}
\includegraphics[width=0.45\textwidth]{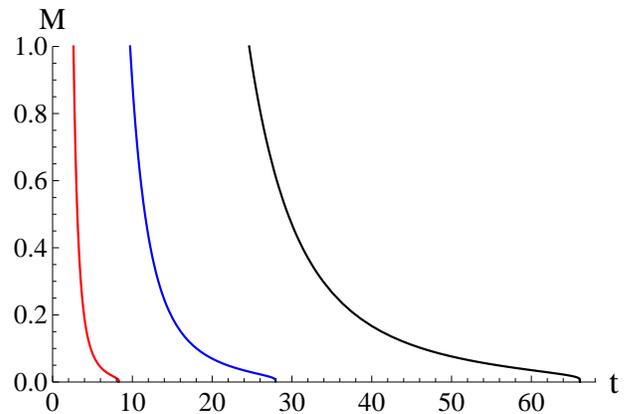}
\caption{The evolution of the black hole in $d=6$ and $k=2$. The initial mass $M_0$ is taken to infinity. From left to right the curves correspond to $\ell=0.05$, $\ell=0.075$ and $\ell=0.1$ respectively.}
\label{fig1}
\end{center}
\end{figure}

In FIG.\ref{fig1} we present some numerical examples of the relationship of black hole mass and time $t$. Here we choose $d=6$, $k=2$, and the initial mass $M_0$ is taken to infinity. From left to right the AdS radius $\ell$ correspond to $\ell=0.05$, $\ell=0.075$ and $\ell=0.1$ respectively. The life time of the black hole also does not diverge. Further analysis shows it remains bounded by a time of the order of $\ell^3$ \cite{Page:2015rxa}.

We also carry out a black hole scan and summarize all theories with a given dimension $d$. We directly apply the numerical method to calculate the life time of the black hole with divergent initial mass. In TABLE \ref{tab1} all the possible cases up to 11 dimensions, which is conjectured to be the arena for the ultimate unified theory, are presented. The life time of the black hole is in the order of $\ell^{d-2k+1}$, which is consistent with the result of the analytic calculation.

\begin{table}[!htbp]
\centering
\begin{tabular}{|c|c|c|c|c|c|c|c|c|c|}
\hline
\backslashbox{$k$}{$d$}{} ~&~  $4$ ~&~ $5$ ~&~ $6$ ~&~ $7$ ~&~ $8$ ~&~ $9$ ~&~ $10$ ~&~ $11$\\
\hline
$1$ ~&~ $\ell^3$ ~&~ $\ell^4$ ~&~ $\ell^5$ ~&~ $\ell^6$ ~&~ $\ell^7$ ~&~ $\ell^8$ ~&~ $\ell^9$ ~&~ $\ell^{10}$ \\
\hline
$2$ ~&~ - ~&~ - ~&~ $\ell^3$ ~&~ $\ell^4$ ~&~ $\ell^5$ ~&~ $\ell^6$ ~&~ $\ell^7$ ~&~ $\ell^8$ \\
\hline
$3$ ~&~ - ~&~ - ~&~ - ~&~ - ~&~ $\ell^3$ ~&~ $\ell^4$ ~&~ $\ell^5$ ~&~ $\ell^6$ \\
\hline
$4$ ~&~ - ~&~ - ~&~ - ~&~ - ~&~ - ~&~ - ~&~ $\ell^3$ ~&~ $\ell^4$ \\
\hline
\end{tabular}
\caption{The life time of the black hole with divergent initial mass in different values of $d$ and $k$.}
\label{tab1}
\end{table}

\subsection{The case of $d=2k+1$}

\begin{figure}
\begin{center}
\includegraphics[width=0.45\textwidth]{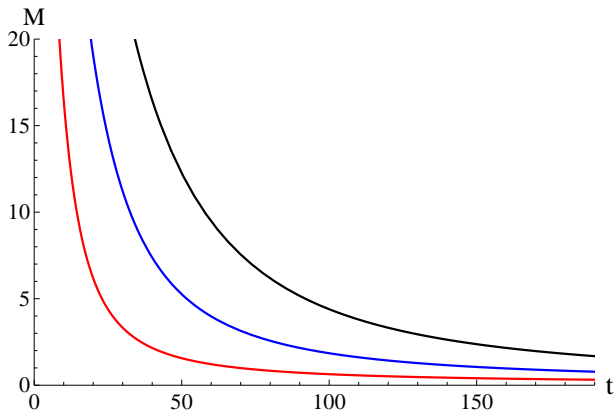}
\caption{The evolution of the black hole in $d=5$ and $k=2$. From left to right the curves correspond to $\ell=0.05$, $\ell=0.075$ and $\ell=0.1$ respectively.}
\label{fig2}
\end{center}
\end{figure}

In the case of $d=2k+1$, the $f(r)$ takes the form
\begin{equation}
f(r)=1+\frac{r^2}{\ell^2}-(2G_k M+1)^{\frac{1}{k}},
\end{equation}
and the black hole temperature reads
\begin{equation}
T=\frac{r_+}{2\pi \ell^2}.
\end{equation}
We can find $\frac{f(r)}{r^2}$ is a monotonically increasing function and approaches $\frac{1}{\ell^2}$ as $r_+\rightarrow \infty$, thus the impact factor $b_c=\ell$. Inserting $b_c=\ell$, $T=\frac{r_+}{2\pi \ell^2}$, and black hole mass into \eqref{law}, we have
\begin{align}
\mathrm{d}t=-kr_+ \big(1+\frac{r_+^2}{\ell^2}\big)^{k-1}\big(\frac{2\pi}{r_+}\big)^{2k+1}\mathrm{d}r_+,
\end{align}
Integrating the above formula from any initial black hole radius to $0$, we can find the life time of the black hole is always \emph{divergent}. In FIG.\ref{fig2} we present some numerical examples in $d=5$ and $k=2$. From left to right the curves correspond to $\ell=0.05$, $\ell=0.075$ and $\ell=0.1$ respectively. In contrast with the the higher dimensional cases, when the black hole evaporates, the temperature also decreases so the evaporation process is increasingly difficult. Small black holes are stable against the decay by Hawking radiation. The black hole will take infinite time to evaporate away, independently from the initial black hole states. This obeys the third law of black hole thermodynamics.

There are two things worth highlighting in this case. The first one is the final state, which corresponds to $M\rightarrow 0$ and $T\rightarrow 0$, is not the true vacuum state. The pure AdS space-time, corresponding to $M=-\frac{1}{2G_k}$, cannot be reached just by black hole evaporation process. The second one is the difference from conformal (Weyl) gravity \cite{Xu:2017ahm,Xu:2018liy}, where a final state with vanishing temperature that cannot be achieved in finite time also exists. However, in conformal (Weyl) gravity the final state corresponds to an extremal black hole with finite size, while in our case it corresponds to $r_+\rightarrow 0$.

\subsection{Asymptotically flat limit $(\ell\rightarrow \infty)$}

In this case the form of $f(r)$ reads
\begin{equation}
f(r)=1-\bigg(\frac{2G_{k}M}{r^{d-2k-1}}\bigg)^{1/k},
\end{equation}
and the formula of the photon orbit $r_p$ remains the same with the finite $\ell$ cases, which is
\begin{equation}
r_p=\bigg(2M\Big(\frac{d-1}{2k} \Big)^k\bigg)^{\frac{1}{d-2k-1}}.
\end{equation}
The impact factor $b_c$ can be obtained as
\begin{equation}
b_c=\frac{r_p}{\sqrt{f(r_p)}}=\bigg(\frac{d-1}{d-2k-1}\bigg)^{\frac{1}{2}}\bigg(2M\Big(\frac{d-1}{2k} \Big)^k\bigg)^{\frac{1}{d-2k-1}},
\end{equation}
so we have
\begin{equation}
b_c\sim M^{\frac{1}{d-2k-1}}.
\end{equation}
Similarly, the black hole temperature
\begin{equation}
T=\frac{(d-2k-1)}{4\pi \kappa _{B}k}\frac{1}{r_{+}}\sim M^{-\frac{1}{d-2k-1}}.
\end{equation}
Thus, the Hawking emission power
\begin{eqnarray}
\frac{\mathrm{d} M}{\mathrm{d}t}=-C b_c^{d-2} T^d \sim -M^{-\frac{2}{d-2k-1}}
\end{eqnarray}
Integrating the above formula for initial black hole with mass $M_0$, we can find the life time of the black hole will be the order of
\begin{equation}
t\sim M_0^{\frac{d-2k+1}{d-2k-1}}.
\end{equation}
In $d=4$ and $k=1$, we recover the well-known result $t\sim M_0^3$ in Einstein gravity.

\begin{figure}
\begin{center}
\includegraphics[width=0.45\textwidth]{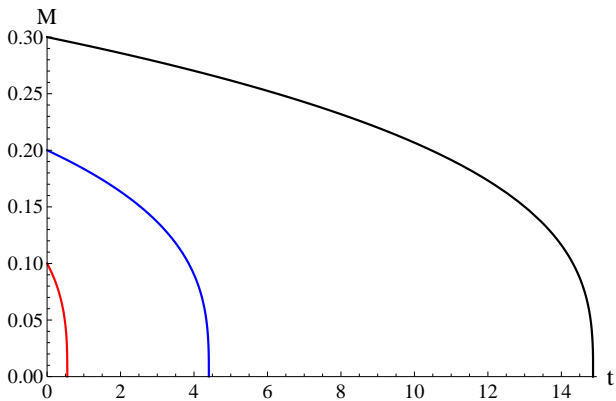}
\caption{The evolution of the black hole when $d=6$ and $k=2$ in asymptotically flat limit $l\rightarrow \infty$. From bottom to top the initial black hole mass $M_0$ corresponds to $M_0=0.1$,$M_0=0.2$ and $M_0=0.3$.}
\label{fig3}
\end{center}
\end{figure}

We present some numerical examples of $d=6$ and $k=2$ in FIG.\ref{fig3}. From bottom to top the curves correspond to $M_0=0.1$, $M_0=0.2$, $M_0=0.3$ respectively. Using the above formula we can see the life time of the black hole satisfies $t\sim M_0^3$, which grows as the cube of the initial mass.

\section{Summary and discussion}
In the present work we explore the black hole evaporation process in Lovelock gravity with diverse dimensions. The action of the Lovelock gravity contains a sum of dimensionally-extended Euler densities, and the equations of motion remain second order. By selecting the appropriate coefficients, the space-time solutions can possess a unique AdS vacuum with a fixed cosmological constant. We divide the Lovelock gravity in AdS space-time into two cases: $d>2k+1$ and $d=2k+1$. In the case of $d>2k+1$, the black hole is in an analogy with the Schwarzschild AdS black hole. The life time of black hole does not diverge with any finite $l$, and it remains bounded by a time of the order of $l^{d-2k+1}$. We can recover Page's result on the Einstein gravity in $k=1$. In the case of $d=2k+1$, the black hole resembles the three dimensional black hole. The space-time solutions predict the existence of a mass gap separating the pure AdS space-time and black hole vacuum. The black hole vacuum corresponds to $T=0$, so the black hole will take infinite time to evaporate away, independently from the initial states, which obeys the third law of black hole thermodynamics. We also consider the asymptotically flat limit by taking $l\rightarrow \infty$. The only non-vanishing term in the equation of motion is the $k$-th one, and the black hole solutions only exist in $d>2k+1$. For an initial mass $M_0$, the life time of the black hole is in the order of $t\sim M_0^{\frac{d-2k+1}{d-2k-1}}$.

Similar analysis may also be extended to Lovelock black holes with electric charge, rotation, and topologically nontrivial AdS asymptotics. In the present work we only consider the asymptotically AdS and flat space-time, while physics in de Sitter space-time has remained so far open problems \cite{Andriot:2019wrs}. It would be interesting to investigate black hole evaporation process in de Sitter space-time (see, e.g. \cite{Kanti:2005ja,Pappas:2016ovo}). Furthermore, we apply the geometrical optics approximation and assume the emitted massless quanta move along null geodesics. However, to get a precise lifetime one has to compute the grey body factors of all the particles \cite{Gubser:1997qr,Mathur:1997et,Klemm:1998bb,Kanti:2002nr,Kanti:2002ge,
Harris:2003eg,Kanti:2004nr,Grain:2005my,Cardoso:2005vb,Cardoso:2005mh,Harmark:2007jy,Boonserm:2008zg,Konoplya:2010vz,Chen:2010ru,Miao:2017jtr,
Zhang:2017yfu,Konoplya:2019hml,Konoplya:2019ppy}. On the other hand, near the end of the evaporation, effective field theory starts to fail, and quantum gravity effect may stop the evaporation, leaving a black hole ``remnant'' \cite{Adler:2001vs,Chen:2014jwq,Ong:2018syk,Yao:2018ceg}. We hope to be able to consider these questions in our future work.

\begin{acknowledgments}
We acknowledge the support by the National Natural Science Foundation of China (No.11875160), the NSFC Guangdong Joint Fund (U1801661), Natural Science Foundation of Guangdong Province (2017B030308003), the Guangdong Innovative and Entrepreneurial Research Team Program (No.2016ZT06D348), and the Science Technology and Innovation Commission of Shenzhen Municipality (ZDSYS20170303165926217, JCYJ20170412152620376, JCYJ20170817105046702).
\end{acknowledgments}

\providecommand{\href}[2]{#2}\begingroup\raggedright\endgroup

\end{document}